# Numerical Integration of Slater Basis Functions Over Prolate Spheroidal Grids

Alexander Stark[1], Nathan Meier[1], Jeffrey Hatch[1], Joshua Kammeraad[1], Duy-Khoi Dang[1], Paul Zimmerman[1*]

[1]Department of Chemistry, University of Michigan, Ann Arbor, Michigan, US.
*paulzim@umich.edu

**Abstract**

Slater basis functions have desirable properties that can improve electronic structure simulations, but improved numerical integration methods are needed. This work builds upon the SlaterGPU library for evaluation of Hamiltonian matrix elements in the resolution-of-the-identity approximation. In particular, a Prolate Spheroidal grid will provide sufficient integral accuracy to employ larger basis sets (quadruple-zeta and greater) in practical computations involving polyatomics. To integrate 3-center Coulomb and nuclear attraction terms, an improved grid representation around the 3rd center is introduced. The RMSEs of the integral quantities are evaluated and compared to the previous numerical integration method used in SlaterGPU (Becke Partitioning), resulting in a ~3 order of magnitude reduction in the error for 2-center integral quantities. The procedure is generally applicable to polyatomic systems, GPU accelerated for high performance computing, and tested on self-consistent field and full configuration interaction wavefunctions. Results for a number of 3-atom models as well as propanediyl ($C_3H_6$) demonstrate the reliability of the new integration scheme.

## 1. Introduction

Electronic structure theory can provide insight into a countless number of chemical systems. Practical methods for molecules most often rely on wavefunctions built upon atom-centered, single-electron basis functions. Certain asymptotic properties of exact wavefunctions can be captured in the single-particle basis set,[1] such as cusps at the nuclei that are consistent with the Kato conditions.[2,21] At long distances wavefunctions decay as simple exponentials. These two ideas lead naturally to Slater Type Orbitals (STOs)[1], which have the form

$$\chi_{nlm}(r, \theta, \phi) = N r^{n-1} e^{-\xi r} Y^{lm}(\theta, \phi) \qquad (1)$$

$N$ is a normalization factor, $\xi$ is the exponent defining each basis function, $n, l, m$ are the principal quantum numbers, and $Y^{lm}$ are spherical harmonics.[2-3] STOs are known to provide accurate descriptions of polarizability, intermolecular interactions, and nuclear shielding, as these properties are sensitive to cusp or decay of the wavefunction.[4-6] The widespread use of Slater functions as electronic structure basis sets, however, has been significantly hindered due to the requirement for numerical integration.

While STOs correctly describe physical properties, their features—such as steepness near the nucleus, long tails, and potentially high angular momentum components—make them difficult to numerically integrate. Gaussian Type Orbitals (GTOs), in contrast, are analytically integrable[8-11] but cannot precisely capture nuclear cusps or long-range decays. Improved integration protocols could start to bridge the practicality gap between STOs and GTOs, making STOs more readily usable in a wide range of electronic structure theories. This work therefore builds upon our recent efforts[15] to provide highly accurate Slater integrals, especially for their use in wavefunction simulations.

Integration of Slater functions has often been performed using atom-centered integration grids, for instance as done in the Amsterdam Density Functional (ADF) program,[22-23,50-51] and more recently through the SlaterGPU library.[15] These integrals use products of radial and angular grids on each atom, where the grid weights are adjusted to avoid overcounting in regions where the grids are overlapping. While various partitioning methods are possible,[24] Becke's method of fuzzy Voronoi cells[12] is probably the most well-known, due to its widespread use in integration of quantities related to density functional theory (DFT).[31-32] Becke partitioning (BP) in ADF has produced integrals required for the GW approximation.[27-30,33] In SlaterGPU, BP has been used in RKS calculations to produce Kohn-Sham potentials.[26]

Two of the authors recently introduced SlaterGPU,[15] an algorithm for accelerating STO integrations on graphics processing units (GPUs). This GPU library uses parallel numerical integration with efficient vector operators based on mixed-precision arithmetic to keep execution costs low. Equally important to the parallelized code was the use of the Resolution-of-the-Identity (RI)[18-20] approximation to reduce the complexity of Coulomb integrals. Coulomb integration with RI can be performed in 3 dimensions in an STO basis, saving a great amount of computational time compared to 6-dimensional integration. Altogether, the original SlaterGPU implementation provided a practical means to perform electronic structure computations at the Hartree-Fock (HF), Complete Active Space Self-Consistent Field (CASSCF), and Full Configuration Interaction (FCI) levels for double- and triple-zeta STO basis sets. The atomic Becke/Voronoi grids of SlaterGPU, however, are insufficiently precise to treat larger basis sets (e.g., for pentuple-zeta, polarized basis sets).

The quality of numerical integration scheme is tied closely to the coordinate system and quadrature that together form the integration grid.[12] Becke showed that an orthogonal curvilinear coordinate system known as prolate spheroidal (PS) coordinates can be effective for integrals involving two atoms.[13] PS coordinates have an origin between two foci and are composed of a radial part $\mu$ and two angular parts $\nu$ and $\phi$, where $\mu$ is composed of spheroids that encompass these foci, $\nu$ is composed of hyperboloids, and $\phi$ is a plane on the focal axis. An interesting property of PS coordinates is that the $\mu, \nu$ degrees of freedom are closely related to $r_1, r_2$, the distances to the two nuclei that define the coordinate system. Close to the nucleus, $r_1$ and $r_2$ are quadratic in $\mu, \nu$, meaning that Slater functions become Gaussians ($\exp[-\xi r] \to \exp[-\xi a(\mu^2 + \nu^2)]$). This coordinate transformation reduces the steepness of the STOs near the nuclei in the PS coordinate space, without compromising the Slater shape expected by the Hamiltonian. Prolate Spheroidal coordinates will be explained in depth in the Theory section.

While PS coordinates are a natural generalization of spherical coordinates from atoms to diatomics, there is no general procedure to handle cases with more than 2 atoms within this coordinate system. Fortunately, under the RI approximation only 3-center integrals are required, so a complete generalization to many-center integration is not required. A path forward to use PS coordinates in practical STO integrations for molecules is therefore conceivable, as long as careful choice of grid discretization around the third center is made.

The SlaterGPU library and its extension to PS coordinates are designed to generate all of the terms necessary for a non-relativistic electronic Hamiltonian expressed in atom-centered Slater orbitals, e.g. for HF, DFT, and post-self-consistent-field (SCF) methods. The goal of this work is to show that polyatomic systems can be handled within the PS coordinate system, with sufficient accuracy and efficiency to allow for the evaluation of larger basis sets. Not only will this allow increased accuracy in post-SCF correlated methods, it also will be useful in deriving Kohn-Sham orbitals and exchange-correlation potentials to high accuracy.[25-26]

## 2. Theory

### 2.1 Prolate Spheroidal Coordinates

PS coordinates are an orthogonal curvilinear coordinate system in three dimensions ($\mu$, $\nu$, $\phi$). The coordinates are defined with respect to two foci, which in this case will be two atomic positions. If the two foci are defined in Cartesian coordinates at the points $(0,0,a)$ and $(0,0,-a)$ then the relation between Cartesian and PS coordinates is defined by

$$x = a \sinh\mu \sin\nu \cos\phi \quad (2)$$
$$y = a \sinh\mu \sin\nu \sin\phi \quad (3)$$
$$z = a \cosh\mu \cos\nu \quad (4)$$

Any two-center integral can be performed after rotation and translation of this grid, assuming distance $2a$ between the atoms (integration with more than 2 atoms is discussed later on). Radial distances from the two atoms are

$$r_1 = a(\cosh\mu + \cos\nu) \quad (5)$$
$$r_2 = a(\cosh\mu - \cos\nu) \quad (6)$$

As pointed out by Becke[13] for small $r_1$ or $r_2$, a Taylor expansion shows that each of these distances is quadratic in $\mu, \nu$. Slater functions therefore can be efficiently integrated due to the avoidance of the cusp near the nucleus, while maintaining the correct physical cusp shape. The $\nu$ and $\phi$ coordinates span $0 \leq \nu \leq \pi$ and $0 \leq \phi \leq 2\pi$. For a pair of atoms, the $\nu$ coordinate moves between the two nuclei, and the $\phi$ coordinate rotates around the two atoms with cylindrical symmetry. The $\mu$ coordinate spans $0 \leq \mu < \infty$, and therefore is mapped onto a finite range in practice (Figure 1).

The PS grid must be discretized to perform numerical integration, as Slater functions evaluated at points will be used for evaluation. Uniform grids in the $\nu$ and $\phi$ coordinates are plausible, though the $\mu$ coordinate deserves more consideration. For example, the $\mu$ grid points should be more concentrated near the nuclei, where the most rapid changes in basis functions occur. Therefore for $0 < t_1 < 1$ we have

$$\mu = C_1 \tanh^{-1}(t_1) \quad (7)$$

where $C_1$ and the maximum value of $t_1$ together fix the maximum value of $\mu$. The $t_1$ coordinates, when uniformly divided, lead to more $\mu$ points near the nuclei. Here, the $\nu$ and $\phi$ grid points are also spaced evenly. Related discretization methods for other coordinate systems can be found in refs 13-16.

Having introduced a method to divide PS coordinate space into discrete volume elements, quadrature within each volume element completes the integration scheme. Due to the way the grid is constructed—where grid discretization places nuclei only at the edges of each volume element—quadrature points will never be evaluated at a nucleus. This can be done using Gauss-Legendre quadrature for all dimensions, which approximates an integral over a volume element without placing points on boundaries.[17] Gauss-Legendre quadrature is exact for polynomials of $2n - 1$ order, allowing rapid convergence with size of quadrature grid.

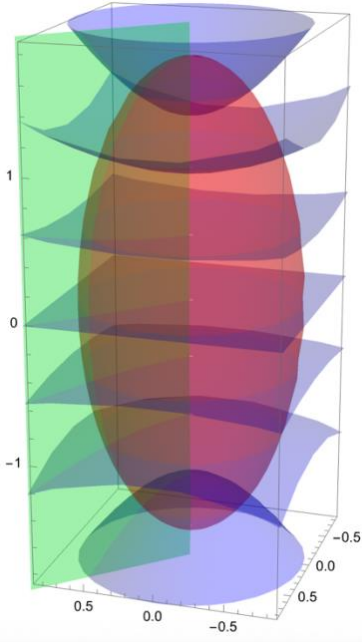

**Figure 1:** The Prolate spheroidal grid on a cartesian grid with spheroids at constant $\mu$ (red), hyperboloids at constant $\nu$ (blue), and planes at constant $\phi$ (green).

A generic integral involving a pair of Slater functions has the form:

$$\int_0^{2\pi}\int_0^{\pi}\int_0^{\infty}\chi_{nlm}(\mu,\nu,\phi)\hat{O}(\mu,\nu,\phi)\chi_{nlm}(\mu,\nu,\phi)d\mu d\nu d\phi \quad (8)$$

Applying discretization and quadrature to this generic integral yields

$$\sum_{i=1}^{M}\sum_{i=j}^{Q}\chi_a(x_{ij})\hat{O}(x_{ij})\chi_b(x_{ij})w(x_{ij}) \quad (9)$$

Where functions will be evaluated at points $x_{ij}$, $\hat{O}$ is the operator of interest, $w(x_{ij})$ is the weight function associated with the quadrature, $Q$ is the number of quadrature points within a volume element, and $M$ is the number of volume elements. The weights are

$$w(x_{ij}) = G(\mu_{ij})G(\nu_{ij})G(\phi_{ij})\Delta V \quad (10)$$

Where $G(\mu_{ij}), G(\nu_{ij}), G(\phi_{ij})$ are the weights from the Gaussian quadrature. The volume element is determined by the spatial extent of the discretized cells.

$$\Delta V = a^3 \sinh \mu_{ij} \sin \nu_{ij} \left(\sinh^2 \mu_{ij} + \sin^2 \nu_{ij}\right)\Delta\mu_{ij}\Delta\nu_{ij}\Delta\phi_{ij} \quad (11)$$

This two-center integration scheme will be shown below to be highly effective.

### 2.2 Treatment of a Third Center

Two-center integration using PS coordinates is efficient since volume elements can be naturally distributed based on the positions of the two centers. Quadrature is also facilitated around these centers due to the grid lines at $\mu = 0$ and $\nu = 0$ or $\pi$. The third center, however, will sit somewhere in an arbitrarily sized volume element, with no particular location relative to the PS grid lines (Figure 2A). To achieve accurate quadrature, the grid lines should be placed to intersect the third nucleus, and the volume elements subdivided. Our grid lines are therefore shifted to accommodate the third center, specifically by moving the nearest volume element borders in $\mu$ and $\nu$ (the position of $\phi$ is trivial, as the three atoms will be placed within the same plane before generating the grid). After the grid lines are moved, the 8 volume elements surrounding the third center are then further divided (Figure 2 B→C). Figure 2 shows a single division surrounding the third center, where 4 cells are divided in half along $\mu$ and $\nu$. Including the $\phi$ degree of freedom, the 8 neighboring cells become 64 cells. This division can be increased to create more cells as necessary to achieve higher precision. The parameter $N_{SP}$ defines how many cells are used to discretize around the third center.

### 2.3 Implementation

The integration grid is set up by enumerating over the discrete volume elements and their weights. First, the angular grid lines are defined for $\nu$ and $\phi$ by dividing $\pi$ and $2\pi$ by $N_\nu$ and $N_\phi$, respectively. The radial component that determines $\mu$ (c.f. equation 12) is divided in the $t_1$ transform uniformly through

$$\Delta t_1 = \frac{1}{N_\mu + 1} \quad (12)$$

Where $N_\mu$ is the number of $\mu$ grid separations. Once the initial 2-center grid is constructed, the 3$^{rd}$ center is considered in the grid. The 3$^{rd}$ center is placed at $\phi = 0$ for alignment within the $\phi$ grid line. Next, the $\mu$ and $\nu$ grid lines closest to the 3$^{rd}$ center are moved to intersect the third center, as shown in Figure 2. Finally, each volume element defined by the above grid lines is divided up via 3-dimensional quadrature to give the full integration grid.

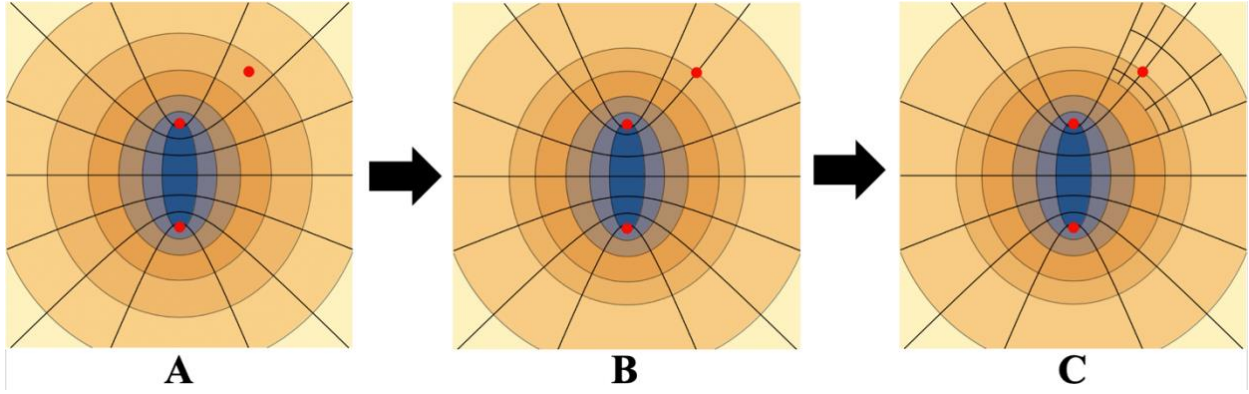

**Figure 2**: The process of reorienting the grid around the third center, A is the initial spacing of grid with the third center being located in one of the volume elements, B moves the $\mu$ and $\nu$ grid lines closest to the third center so that the third center coordinates align with the grid lines, C splits the grid further around the third center for greater accuracy.

**Algorithm 1:** *Coordinate Grid Discretization* : $N_\mu, N_\nu, N_\phi$ are the number of divisions over a coordinate. $\mu$, $\nu$, and $\phi$ represent a point in the center of a volume element and $d\mu$, $d\nu$, and $d\phi$ are the distance from one edge of the volume element to the other edge of the volume element

1   $dx, d\nu, d\phi$ calculated based on user input
2   **for** $i$ *in* $N_\mu$ **do**
3     $t_i, t_{i-1} \leftarrow dt, i$
4     $\mu_i, \mu_{i-1} \leftarrow t_i, t_{i-1}$
5     $\mu, d\mu \leftarrow \mu_i, \mu_{i-1}$
6     **for** $j$ *in* $N_\nu$ **do**
7       $\nu \leftarrow d\nu, j$
8       **for** $k$ *in* $N_\phi$ **do**
9         $\phi \leftarrow d\phi, k$
10        Save to Grid$[\mu, \nu, \phi, d\mu, d\nu, d\phi]$

**Algorithm 2:** *Quadrature Grid Initialization* : $Q_\mu, Q_\nu, Q_\phi$ are the quadrature points of a respective coordinate and $w_\mu, w_\nu, w_\phi$ are the associated weights

1   **for** $n$ *in volume elements* **do**
2     Read in volume element $\mu_n, \nu_n, \phi_n, d\mu_n, d\nu_n, d\phi_n$
3     **for** $i$ *in* $Q_\mu$ *points* **do**
4       $\mu_{ni} \leftarrow Q_{i\mu}, \mu_n, d\mu_n$
5       Read in $w_{i\mu}$
6       **for** $j$ *in* $Q_\nu$ *points* **do**
7         $\nu_{nj} \leftarrow Q_{j\mu}, \mu_n, d\mu_n$
8         Read in $w_{j\nu}$
9         **for** $k$ *in* $Q_\phi$ *points* **do**
10           $\phi_{nk} \leftarrow Q_{k\mu}, \mu_n, d\mu_n$
11           Read in $w_{k\phi}$
12           $Qpoints[x, y, z] \leftarrow \mu_{ni}, \nu_{nj}, \phi_{nk}$
13           $weight[w] \leftarrow w_{i\mu}, w_{j\nu}, w_{k\phi}$

## 3. Computational Details

Most results in this work utilize an all-electron triple-zeta basis set with polarization functions (denoted TZ), while others utilize quadruple- and pentuple-zeta basis sets (denoted QZ and 5Z, respectively). Basis sets were constructed to be even tempered (with exponents $\zeta = \alpha\beta^n$, $n = 0, -1, \cdots, -N$) where N depends on the angular momentum and row on the periodic table. Auxiliary functions were generated in a combinatorial fashion from the original basis by adding together the angular momentum $\ell$ as well as the exponents of all pairs of functions on each atom. For each channel $\ell$, the minimum and maximum summed exponents ($\zeta_{max}$ and $\zeta_{min}$) were selected to define the range for the auxiliary basis. The auxiliary basis is then generated through an even-tempered procedure with $\zeta_{max}$ and $\zeta_{min}$ as its limits.[60-63] Finally, all $m$ degrees of freedom for each $\ell$ were enumerated. The basis sets are provided in the supporting information (Table S1).

Figures 3, 4, 5, 6 and Table 1 all analyze integral matrices directly. Figures 3, 4, 5 and Table 1 do so through analysis of the Root Mean Square Error (RMSE) which is defined by

$$RMSE = \sqrt{\frac{\sum_{i=1}^{n}(g_i^{ref} - g_i)^2}{n}} \quad (13)$$

$g_i$ are matrix elements in the integral matrices, $g^{ref}$ is a matrix produced from an accurate calculation with a large grid, and g is a matrix produced from some less accurate calculation.

Coulomb integrals are computed under the RI approximation, where 4-index integrals are determined as follows

$$(ij|kl) \approx \sum_{PQ}(ij|P)(PQ)^{-1}(Q|kl) \quad (14)$$

These two 3-index terms can be expressed as

$$(P|ij) = \iint \chi_P(r_1)\frac{1}{r_{12}}\chi_i(r_2)\chi_j(r_2)dr_1 dr_2 \quad (15)$$

and simplified to

$$(P|ij) = \int V_C^P(r)\chi_i(r)\chi_j(r)dr \qquad (16)$$

All CI computations in Table 3 were run in a neutral state (singlet spin for all, except doublet spin for NO, N₃, and NO₂). The geometries for the diatomic systems in the table were obtained from the Computational Chemistry Comparison and Benchmark DataBase (https://cccbdb.nist.gov/). Triatomic geometries were optimized using Q-Chem version 5.2,[59] geometries for specific species and information about methods used to obtain these geometries can be found in the SI (Table S2).

HBCI, used in Tables 3 and 4, is a select-CI approach that returns a close approximation to the full CI limit.[45-48] The energy thresholds $\varepsilon_1$ and $\varepsilon_2$ control the extent of recovery of correlation through variational ($\varepsilon_1$) and perturbative ($\varepsilon_2$) steps. In Table 3 HBCI was performed using $\varepsilon_1 = 1 \times 10^{-4}$ Ha and $\varepsilon_2 = 1 \times 10^{-7}$ Ha except in cases which possess greater than 15 valence electrons (N₂O, NO₂, OF₂, SO₂, CS₂), which used $\varepsilon_1 = 5 \times 10^{-4}$ Ha and $\varepsilon_2 = 5 \times 10^{-7}$ Ha. The iFCI method, used in Figure 7, truncates the search for configurations in the Hilbert space further by defining localized molecular orbitals as base units for correlation. A many-body expansion combines these units to systematically recover correlation from a reference state (a valence bond, perfect-pairing wave function), ensuring convergence to full CI as the expansion level, $n$, is increased.[64-66] This allows for polynomial scaling of the iFCI method while maintaining a similar accuracy to HBCI. Here, $n = 3$ recovers the majority of the correlation energy and simplifies the computation of the 24 electron in 225 orbital all electron (core + valence) space. See refs 64, 71 and 72 for further details of this approach. The HBCI solver in the iFCI calculation uses energy thresholds of $\varepsilon_{1,doubles} = 5 \times 10^{-4}$ Ha, $\varepsilon_{1,singles} = 2.5 \times 10^{-4}$, and $\varepsilon_2 = 1 \times 10^{-7}$ Ha. The frozen core approximation has been used in all calculations, besides in Table 4 and Figure 7, where all electrons were correlated. The CCSD(T) method[73,74] was also performed in Figure 7, applied to the same 1,3 propanediyl system, core correlation was included and a UHF reference was utilized.

For comparisons to GTOs for the methylene and 1,3 propanediyl systems, the cc-pVXZ family was used, the cc-pVXZ-RIFIT auxiliary basis was utilized in calculations on methylene, no auxiliary basis was used for 1,3 propanediyl. Here X is T, Q, and 5 for polarized triple-, quadruple-, and pentuple-zeta basis sets, respectively. Heat-bath configuration interaction (HBCI)[48] was used as a representative electronic structure method, which closely approximates the full CI energy. The Nvidia HPC SDK 25.5 compiler suite with OpenACC was used to compile SlaterGPU and HBCI. The calculations using GTOs on the CH₂ system in Table 4 were run using Q-Chem version 5.2.[59] The CCSD(T) calculations used in Figure 7 were run using ORCA 6.0.1.[67-70]

When discretizing the initial PS grid, the scalar $C_1$ transforms the overall grid size, depending on the spacing between the atoms. This is set to

$$C_1 = 2.3a^{1/4} \qquad (17)$$

The angular components are left untransformed.

There are 5 variables ($N_\mu$, $N_\nu$, $N_\phi$, quadrature points (Q), and third-center split ($N_{SP}$)) which need to be chosen to specify the grid for PS integration. To demonstrate convergence with respect to grid size, we vary the radial grid size as well as the angular grid size but fix $Q = 4$ and $N_{SP} = 3$ unless otherwise mentioned. Table S3 shows the choice of grid sizes for the radial and angular grids used in Figures 3, 4 and 5. For comparisons in Table 1 between the BP integration (with a grid composed of 5810 angular points and 120 radial points per atom) and the similarly sized PS grid ($N_\mu$: 26 $N_\nu$: 32 $N_\phi$: 14 Q: 4 $N_{SP}$: 3), these results are compared to a large PS grid ($N_\mu$: 80 $N_\nu$: 70 $N_\phi$: 50 Q: 4 $N_{SP}$: 3 ). Table 2 uses the same BP grid as Table 1. Figure 6 as well as Tables 2, 3 and 4 use the same large PS grid as Table 1. The iFCI calculations on the 1,3 propanediyl system in Figure 7 use a PS grid with the following parameters $N_\mu$: 80 $N_\nu$: 60 $N_\phi$: 40 Q: 4 $N_{SP}$: 4.

## 4. Results and Discussion

As a starting point, two molecules were selected to demonstrate and benchmark the new PS integration method: ClF and OF₂. The 2- and 3-center integrals needed for the electronic Hamiltonian under the RI approximation were computed using a range of grid sizes. By varying the number of radial and angular discretization points, the convergence and numerical accuracy of PS integration will be discussed and then compared to integrals from the original SlaterGPU method.[15] The Slater basis for the initial tests is of triple-$\zeta$ quality, including s, p, and d angular momentum functions and exponents ranging from 0.75 to 22.0. The auxiliary basis contains up to g functions.

Figures 3 and 4 show that the PS integration technique produces low errors and smooth convergence with respect to radial and angular discretization for all Hamiltonian elements for ClF. The most challenging cases are the electron-nuclear attraction integrals, due to the singularity at each nucleus. Regardless, the smallest integration grid for 2-center Coulomb (Figure 3) reaches an RMSE of order $10^{-7}$, and increased grid discretization lowers errors to order $10^{-12}$ RMSE. These errors are lower than those of the original SlaterGPU method (a detailed comparison is given later on). The remaining integrals—the overlap, kinetic, electron nuclear and 3-center Coulomb integrals—all show excellent convergence. Obtaining RMSE below $10^{-10}$ does not require especially large grids. The largest grids investigated here show RMSEs around $10^{-13} - 10^{-14}$, which is close to what is possible with double precision arithmetic. These results

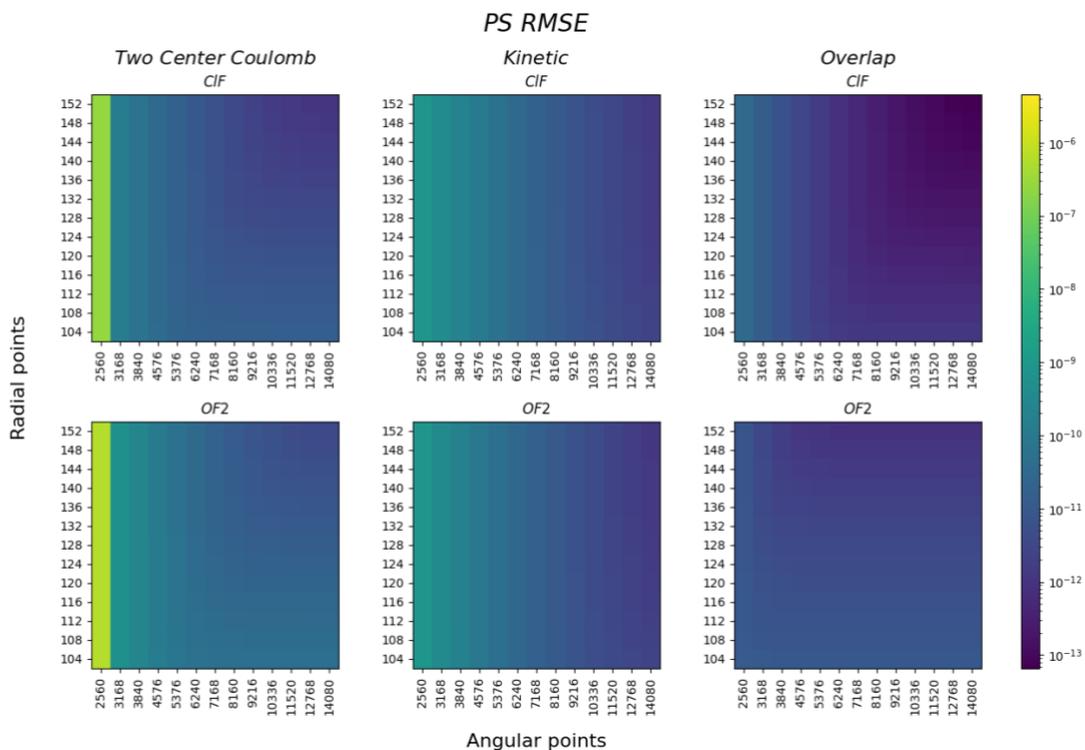

**Figure 3:** Overlap, kinetic energy, and 2 center Coulomb repulsion RMSE (Ha) analyzed for ClF and $OF_2$ using the PS method in a TZ basis.

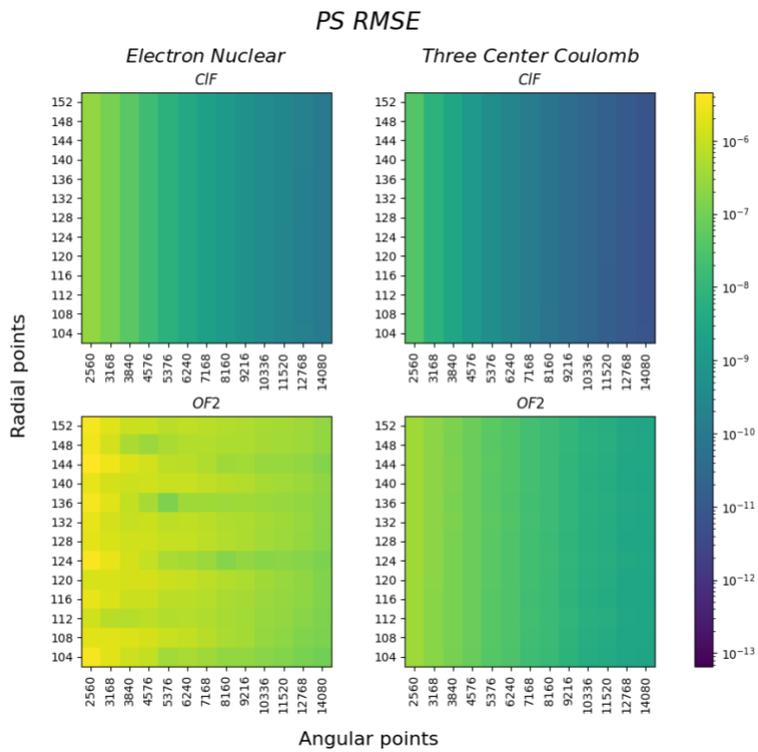

**Figure 4:** Electron-nuclear attraction and 3-center Coulomb repulsion RMSE (Ha) analyzed for ClF and $OF_2$ using the PS method in a TZ basis. The grid selection and legend are the same as in Figure 3.

confirm two-atom integrals are well matched to numerical PS integration.[36,37]

For the three atom $OF_2$ system, the Coulomb and the electron-nuclear attraction integrals are somewhat more challenging for PS integration. This can be seen in comparison to the 2-atom integrals for ClF in Figure 4, revealing decreased accuracy for the 3-atom integrals. Increasing the grid sizes can systematically lower the RMSE, bringing the errors from the smallest to the largest grid from $10^{-6}$ to $10^{-8}$ for electron-nuclear attraction, and $10^{-7}$ to $10^{-9}$ for the 3-center Coulomb. As will be shown later on, these errors are sufficiently low to allow high-quality wavefunction simulations to be performed. Since the nonrelativistic electronic Hamiltonian under the RI approximation involves only integrals with up to three atoms, these RMSE values are expected to also apply to polyatomic systems. That is, even for a polyatomic system, the grid does not need to be extended beyond 3 centers.

As discussed in section 2.2 and motivated by Figure 4, additional discretization in the PS grid around the third center may be helpful for numerical accuracy. The $N_{SP}$ parameter controls this discretization, so the errors in the electron-nuclear attraction integrals were analyzed for $1 \leq N_{SP} \leq 4$. Figure 5 shows that $N_{SP} = 2$ yields noticeable improvements, but $N_{SP} > 2$ has little utility, at least for the equilibrium geometry of the $OF_2$ molecule. Since the spacing between $\mu$ grid lines grows with $\mu$, the volume elements around the third center grow at large distances. Therefore we expect that higher $N_{SP}$ might have more utility for centers which are farther from each other. To test this hypothesis, a nonequilibrium geometry for $OF_2$ was created by moving one fluorine atom to ~30 times its equilibrium bond distance. $N_{SP} = 3$ significantly improved accuracy over $N_{SP} = 2$, but $N_{SP} = 4$ provided little additional utility. For a molecule with long distances between atoms, the $N_{SP}$ discretization scheme may therefore be useful up to about $N_{SP} = 3$. Quantities supporting these results are given in the SI (Figure S1).

To test the applicability of PS integration on variation in the molecular geometry, a set of paths involving changing nuclear positions were considered. As the geometry changes, the integral values should be smooth and lacking any artifacts from the numerical integration scheme. To do this, two centers are defined: one at the origin and the other displaced by a unit vector in one of the 16 directions. The second center was moved along the unit vector until the distance between the centers was 6 Bohr. Figure 6 illustrates the value of a 2-center Coulomb integral along this path. All integral profiles are smooth, even for difficult cases with high angular momentum, for instance a pair with f and h functions ($\ell = 3$ and $\ell = 5$) on the bottom right corner of Figure 6. A larger list of basis combinations can be found in the SI (Figure S2) that show related results with similar accuracy.

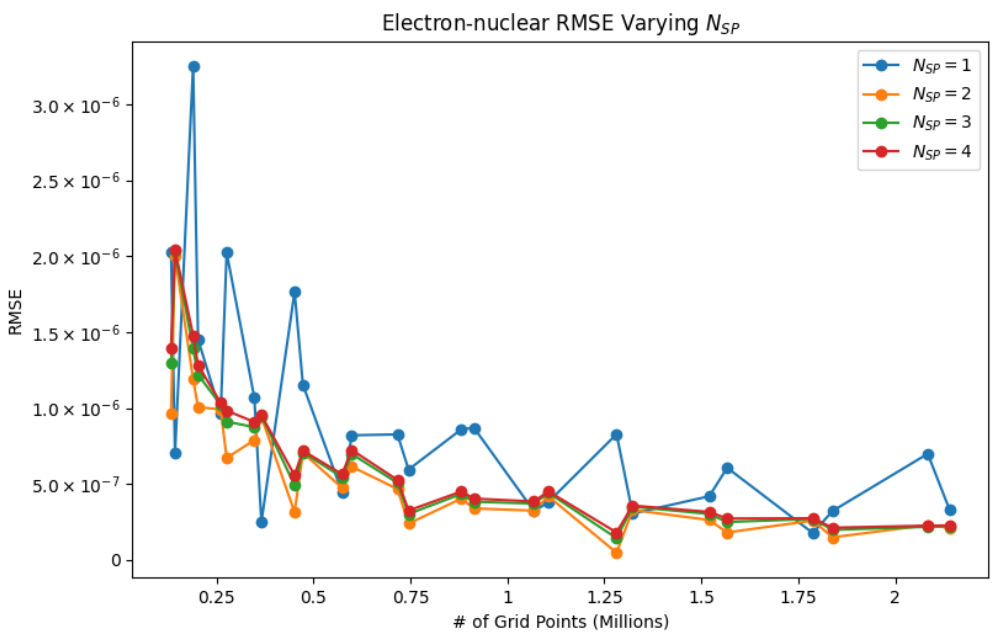

**Figure 5:** Comparison of the RMSE (Ha) of the electron-nuclear attraction elements for different divisions of the third center within the PS integration grid. A single split ($N_{SP} = 2$) gives significant advantages over no splitting, and $N_{SP} > 2$ gives marginal improvements.

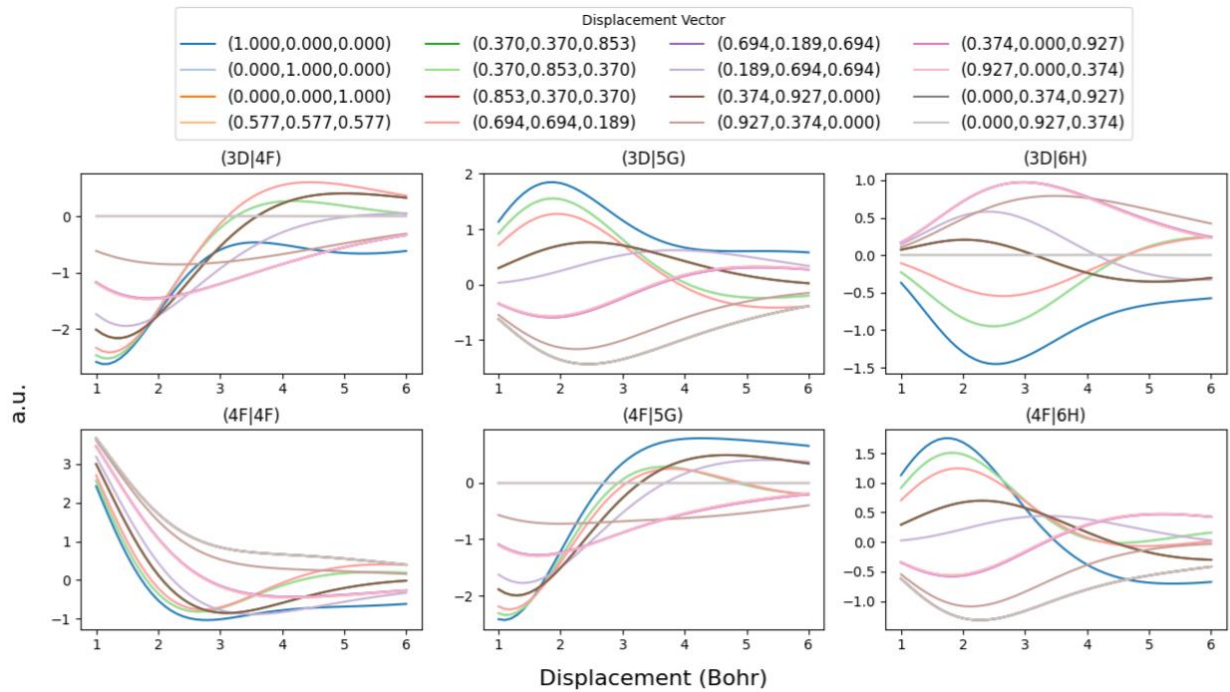

**Figure 6:** Six 2-center Coulomb integrals plotted along 16 vector directions[15] as the two centers are separated. The legend indicates unit vectors for all 16 directions tested, all basis functions used have the exponent $\zeta = 1$, and all basis functions were chosen to have $m = 0$.

**Table 1:** RMSE (Ha) of integrals for $SO_2$ using BP and PS integration with TZ, QZ, and 5Z basis sets. Basis sets TZ, QZ, 5Z are described in the computational details.

| RMSE of integrals on $SO_2$ | | | | | | |
|---|---|---|---|---|---|---|
| | TZ | | QZ | | 5Z | |
| | BP | PS | BP | PS | BP | PS |
| Overlap | $1.3 \times 10^{-08}$ | $8.0 \times 10^{-12}$ | $7.6 \times 10^{-09}$ | $5.4 \times 10^{-12}$ | $1.3 \times 10^{-08}$ | $8.9 \times 10^{-12}$ |
| Kinetic | $4.6 \times 10^{-07}$ | $9.6 \times 10^{-11}$ | $3.1 \times 10^{-07}$ | $3.0 \times 10^{-10}$ | $3.0 \times 10^{-07}$ | $2.8 \times 10^{-10}$ |
| Electron Nuclear | $7.7 \times 10^{-07}$ | $3.0 \times 10^{-07}$ | $5.8 \times 10^{-07}$ | $3.1 \times 10^{-07}$ | $6.0 \times 10^{-07}$ | $3.2 \times 10^{-07}$ |
| 2-Center Coulomb | $3.2 \times 10^{-08}$ | $4.9 \times 10^{-11}$ | $2.0 \times 10^{-08}$ | $6.8 \times 10^{-11}$ | $3.1 \times 10^{-08}$ | $8.1 \times 10^{-11}$ |
| 3-Center Coulomb | $8.7 \times 10^{-09}$ | $6.6 \times 10^{-09}$ | $5.6 \times 10^{-09}$ | $1.3 \times 10^{-08}$ | $7.3 \times 10^{-09}$ | $1.6 \times 10^{-08}$ |

The analysis so far suggests that the PS integration scheme is able to accurately integrate 2- and 3-center quantities of the electronic Hamiltonian. Now, we compare the PS method with the original BP method of the SlaterGPU code. This comparison was performed with relatively similar grid sizes of ~700,000 points. The RMSE obtained for the electron-nuclear and 3-center Coulomb integrals when comparing BP and PS methods are within the same order of magnitude (Table 1). The overlap, kinetic, and 2-center Coulomb integrals have improvements in the error of up to 3 orders of magnitude for PS over BP integration. The improved accuracy of the PS method for 2-center integrals will now be shown to be important for SCF calculations.

The ability to precisely compute integrals with the PS integration method will be a significant advantage to practical electronic structure simulations. Therefore a variety of polyatomic 2- and 3-atom molecules were examined to demonstrate the ability of the PS and BP methods to handle larger basis sets, including basis sets with relatively high angular momentum functions. The TZ, QZ, and 5Z basis sets include two d functions (TZ), three d and one f (QZ), and four d and two f (5Z), respectively. As integral accuracy is lower for BP compared to PS integration, there may be noticeable

**Table 2:** Convergence of Hartree-Fock SCF runs with BP integration. Each number indicates the eigenvalue threshold needed to control numerical stability of the overlap matrix inversion. If the value is 0 (green) all threshold values tested allow for SCF convergence. Higher values indicate instability at this step, due to lower accuracy of the overlap matrix elements. Oscillatory cases (orange) oscillated around the energy obtained through the PS method with a difference less than 0.1 Ha but did not converge. Energies which varied by more than 0.1 Ha are marked in red. In comparison, PS integration converged for all cases with the smallest eigenvalue threshold.

| Eigenvalue Cutoff for SCF Convergence (BP Integration) | | | | | | | | | | | |
|---|---|---|---|---|---|---|---|---|---|---|---|
| Mol | TZ | QZ | 5Z | Mol | TZ | QZ | 5Z | Mol | TZ | QZ | 5Z |
| HF | 0 | 0 | $10^{-5}$ | CS | 0 | $10^{-6}$ | $10^{-6}$ | $N_3$ | 0 | 0 | $10^{-6}$ |
| $C_2$ | 0 | 0 | 0 | SiO | 0 | $10^{-6}$ | 0 | $N_2O$ | 0 | 0 | $10^{-7}$ |
| $N_2$ | 0 | 0 | 0 | ClF | 0 | 0 | 0 | $NO_2$ | 0 | 0 | 0 |
| CO | 0 | 0 | 0 | $Cl_2$ | 0 | 0 | 0 | SFH | 0 | $10^{-7}$ | $10^{-7}$ |
| NO | 0 | 0 | 0 | $H_2O$ | 0 | 0 | 0 | $OF_2$ | 0 | 0 | 0 |
| $O_2$ | 0 | 0 | 0 | HCN | 0 | 0 | 0 | $SO_2$ | 0 | $10^{-7}$ | $10^{-7}$ |
| HCl | 0 | 0 | 0 | OFH | 0 | 0 | 0 | $CS_2$ | 0 | $10^{-7}$ | $10^{-7}$ |

Success — Oscillatory — Far from Convergent

differences in practical electronic structure computations. As a simple test, self-consistent field (SCF) computations at the Hartree-Fock (HF) level were performed. The PS method allows for SCF convergence of all systems in all basis sets explored in Table 2. The BP method, however, performs well for the TZ basis set but is less reliable for the larger basis sets.

To understand why the larger basis sets are much more challenging, we considered the smallest eigenvalues of the overlap matrix for the $SO_2$ system. For the TZ, QZ, and 5Z basis sets, these values are $1.3\times10^{-5}$, $9.8\times10^{-8}$, and $7.1\times10^{-8}$, respectively. The latter two values are close to the limit of the BP integration accuracy, meaning one or more degrees of freedom in the basis may be poorly behaved. The mixed precision used in the evaluation of BP integrals is believed to contribute to the degradation in performance.[15] PS integration, having accuracies of around $10^{-12}$ RMSE, experiences no difficulty through 5Z basis sets.

Further tests of the utility of the PS integrals were done for the same molecules in Table 2, this time using a correlated post-HF method. High-quality wavefunctions were computed using TZ, QZ, and 5Z basis sets at the HBCI level of theory.[45-48] HBCI provides a close approximation to full CI, recovering all static and dynamic correlation available to the basis. Given that all degrees of orbital freedom are accessed by HBCI, errors in the integrals can cause serious problems in the reliability of the CI procedure. The HBCI calculations in Table 3 illustrate the accuracy of the integrals and reference orbitals generated using the PS method. In particular, from TZ to QZ, additional correlation energies of -4.22 mHa/electron on average were found, and from QZ to 5Z, -1.29 mHa/electron, suggesting the relevance of high-quality basis sets for converging these wavefunctions. PS integrals therefore were instrumental in achieving accurate results from large Slater basis sets at a correlated level.

**Table 3:** Heat-bath configuration interaction (HBCI) correlation energies (Ha) for various molecules in TZ, QZ, and 5Z basis sets.

| $E_{corr}$ (Ha) using PS integrals with increasing basis | | | |
|---|---|---|---|
| Mol | TZ | QZ | 5Z |
| HF | -0.2558 | -0.2821 | -0.3005 |
| $C_2$ | -0.3629 | -0.3861 | -0.3895 |
| $N_2$ | -0.3646 | -0.3960 | -0.4091 |
| CO | -0.3477 | -0.3829 | -0.3935 |
| NO | -0.6962 | -0.7309 | -0.7487 |
| $O_2$ | -0.4641 | -0.5138 | -0.5299 |
| HCl | -0.1910 | -0.2373 | -0.2417 |
| CS | -0.3041 | -0.3485 | -0.3526 |
| SiO | -0.3247 | -0.3646 | -0.3738 |
| ClF | -0.4191 | -0.4750 | -0.4948 |
| $Cl_2$ | -0.3490 | -0.4399 | -0.4482 |
| $H_2O$ | -0.2517 | -0.2821 | -0.2912 |
| HCN | -0.3402 | -0.3730 | -0.3811 |
| OFH | -0.4834 | -0.5374 | -0.5632 |
| $N_3$ | -0.9796 | -1.0304 | -1.0537 |
| $N_2O$ | -0.6043 | -0.6643 | -0.6875 |
| $NO_2$ | -1.1706 | -1.2314 | -1.2607 |
| SFH | -0.4117 | -0.4731 | -0.4940 |
| $OF_2$ | -0.7200 | -0.7955 | -0.8395 |
| $SO_2$ | -0.6140 | -0.6965 | -0.7196 |
| $CS_2$ | -0.4646 | -0.5419 | -0.5503 |

Having found that large Slater basis sets can be used in correlated wavefunction computations, we test the PS integration method further on two statically correlated systems. The first case is a well-studied benchmark[38-44] for electronic structure methods: the singlet-triplet spin gap of methylene ($CH_2$). The $CH_2$ transition between $^1A_1$ and $^3B_1$ states were obtained via extrapolation of a series of HBCI computations to the FCI limit with the following HBCI convergence parameters $\varepsilon_1 = 2 \times 10^{-4}, 1 \times 10^{-4}, 0.5 \times 10^{-4}$ Ha and $\varepsilon_2 = 5 \times 10^{-7}$ Ha. Further details are available in SI (Figures S3 and S4). The HBCI/STO results obtained are close to Diffusion Monte Carlo (DMC) and internally contracted multireference configuration interaction (CMRCI+Q). Both the HBCI/STO and HBCI/GTO calculations behave well and come within 0.36-0.72 kcal/mol and 0.15-0.51 kcal/mol of the experimental spin gap (Table 4).

PS integrals via SlaterGPU are applicable to molecular systems with more than 3 atoms, as the electronic Hamiltonian can be constructed by computing integrals over subsets of 3 atoms at a time. As an example, incremental FCI (iFCI)[64] computations were run on 1,3 propanediyl to obtain the vertical singlet-triplet gap. iFCI is able to recover impressive amounts of correlation energy at polynomial scaling: all core and valence electrons were correlated within a 225 orbital active space (full CI). When comparing the singlet-triplet gap energies of the iFCI method with CCSD(T) we observe that the predicted energies for the vertical excitation differ by more than 1 kcal/mol. The predicted energy difference can be explained by the use of an unrestricted reference for CCSD(T) as spin contamination can be a problem for open-shell systems,[75,76] in this case likely causing a cancellation of errors. Based on the natural orbitals obtained from the iFCI calculation, the 1,3 propanediyl system has two radical centers located on carbons 1 and 3. Simultaneously, the natural orbital occupation numbers in the ground singlet state indicate the system has significant biradicaloid character (Figure 7).

**Table 4:** Two states of $CH_2$ ($^1A_1$ and $^3B_1$), at various correlated levels of theory. Geometries for $^1A_1$ and $^3B_1$ $CH_2$ were obtained from ref. 34. The bases used for the GTO calculations are the cc-pVXZ basis sets while the STO bases are the same as referenced in Table 2. The energy reported under CBS was obtained in the Complete Basis Set limit. Units are in kcal/mol.

| Method | TZ | QZ | 5Z | $CBS^f$ | Expt. | Basis |
|---|---|---|---|---|---|---|
| HBCI-PS | 11.1 | 9.93 | 9.72 | | | Slater |
| HBCI-GTO | 10.5 | 9.71 | 9.51 | | | Slater |
| Expt.[a] | | | | | 9.0 | |
| Expt.[b] | | | | | 9.36 | |
| SF-CIS[c] | 20.4 | | | | | Gaussian |
| SF-CIS(D)[c] | 14.1 | | | | | Gaussian |
| EOM-SF-CCSD(dT)[d] | | 9.7 | | | | Gaussian |
| VMC[e] | 9.92 | | | | | Slater |
| DMC[e] | 9.36 | | | | | Slater |
| CMRCI+Q[f] | | | | 8.97 | | Gaussian |
| FCI[g] | 11.1 | | | | | Gaussian |

[a] Ref. 38
[b] Ref. 39
[c] cc-pVTZ, all-electrons correlated Ref. 40
[d] aug-cc-pVQZ, all-electrons correlated Ref. 41
[e] TZ2P/Jastrow, all-electrons correlated Ref. 42
[f] Complete basis set limit, all-electrons correlated Ref. 43
[g] TZ2P Frozen-core Ref. 44

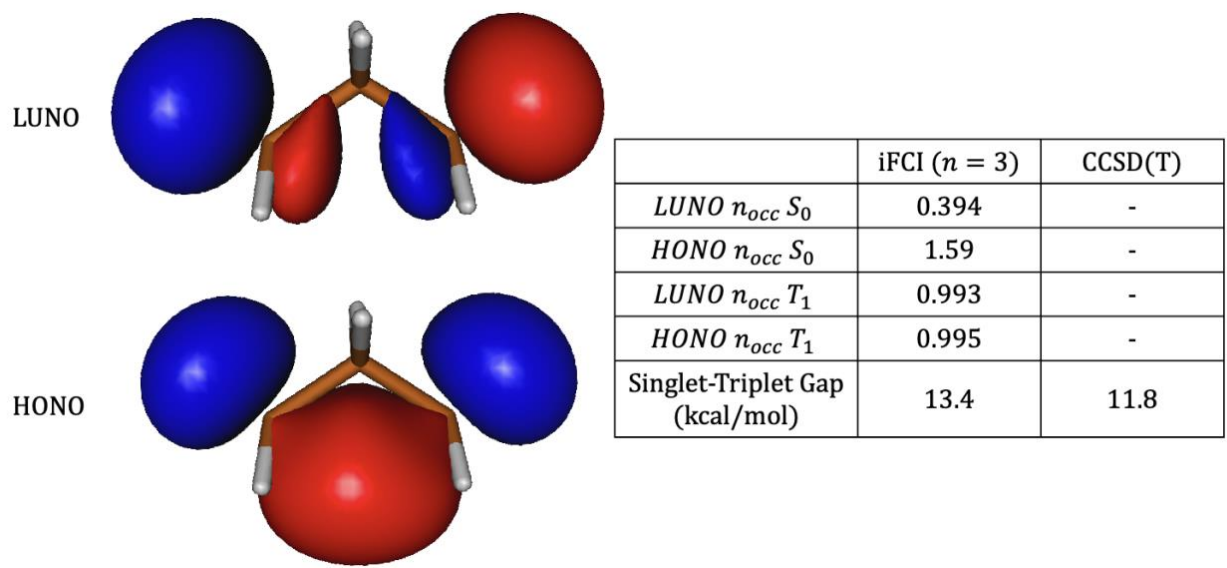

**Figure 7:** HONO and LUNO orbitals and occupation numbers for the singlet and triplet states of 1,3 propanediyl as well as the singlet-triplet gap for the vertical excitation. iFCI uses the STO QZ basis while CCSD(T) uses a GTO basis, cc-pVQZ.

## 5. Conclusion

Leveraging the advancements made in the original SlaterGPU paper,[15] this work reports an efficient numerical integration scheme to allow for high-accuracy construction of electronic structure matrix elements. The two-center PS integrals are around 3 orders of magnitude more accurate than the prior integration scheme (see Table 1), with the 3-center integrals being of similar accuracy. The increased accuracy of the overlap integrals in particular allows for the use of larger basis sets in practical electronic structure simulations, since the integrals generated via BP are inadequate for the convergence of QZ and 5Z basis SCF simulations (c.f. Table 2). Examples of high precision CI computations demonstrate the practical utility of the new scheme (Table 3, Table 4, Figure 7).

The method presented excels in calculations which require an accurate description of the cusp and exponential wavefunction tails. This is notably beneficial in inverse DFT applications where the cusp conditions can greatly impact the exchange-correlation potential.[25-26] The Prolate Spheroidal scheme should allow for less numerical artifacts and improve the output obtained from the inverse Kohn-Sham problem.[51-58]

**Supporting Information Available**

The Supporting information contains

Basis set zeta values for all centers used in this paper, Geometries for all systems, Grid sizes used to generate Figures 3, 4, and 5. Plot of $N_{SP}$ convergence for the nonequilibrium $OF_2$ geometry. Plots of all basis combinations for 2-center Coulomb integrals varying with displacement along 16 vectors. Plots which show extrapolation to FCI for Table 4 in GTO and STO basis sets.

**Acknowledgements**

This work was supported by the U.S. Department of Energy, Office of Science, Basic Energy Sciences under Award DE-SC0022241. This work was also supported by the National Energy Research Scientific Computing Center for computing resources. The authors wish to thank David Braun for continued computational support.

| | iFCI ($n = 3$) | CCSD(T) |
|---|---|---|
| LUNO $n_{occ}$ $S_0$ | 0.394 | - |
| HONO $n_{occ}$ $S_0$ | 1.59 | - |
| LUNO $n_{occ}$ $T_1$ | 0.993 | - |
| HONO $n_{occ}$ $T_1$ | 0.995 | - |
| Singlet-Triplet Gap (kcal/mol) | 13.4 | 11.8 |

# Supplemental Information

# Numerical Integration of Slater Basis Functions Over Prolate Spheroidal Grids

Alexander Stark[1], Nathan Meier[1], Jeffrey Hatch[1], Joshua Kammeraad[1], Duy-Khoi Dang[1], Paul Zimmerman[1]

[1]Department of Chemistry, University of Michigan, Ann Arbor, Michigan, US.
*paulzim@umich.edu

## Table of Contents

## I. Slater Basis Sets

**Table S1:** Slater basis exponents for TZ, QZ, and 5Z basis functions applied to molecules in Figures 3, 4, 5, 7, and S1 and Tables 1, 2, 3 and 4.

| TZ | | QZ | | 5Z | |
|---|---|---|---|---|---|
| **H** | | | | | |
| 1s | 2.280 | 1s | 4.000 | 1s | 4.100 |
| 1s | 1.520 | 1s | 2.500 | 1s | 2.929 |
| 1s | 1.013 | 1s | 1.5625 | 1s | 2.092 |
| 2p | 2.100 | 1s | 0.977 | 1s | 1.494 |
| 2p | 1.500 | 2p | 3.200 | 1s | 1.067 |
| | | 2p | 2.286 | 2p | 4.000 |
| | | 2p | 1.633 | 2p | 2.353 |
| | | 3d | 2.200 | 2p | 1.384 |
| | | | | 2p | 0.814 |
| | | | | 3d | 3.400 |
| | | | | 3d | 1.700 |
| **C** | | | | | |
| 1s | 8.689 | 1s | 11.271 | 1s | 15.000 |
| 1s | 5.323 | 1s | 7.653 | 1s | 10.714 |
| 1s | 3.261 | 1s | 5.196 | 1s | 7.653 |
| 1s | 1.998 | 1s | 3.528 | 1s | 5.466 |
| 1s | 1.224 | 1s | 2.396 | 1s | 3.905 |
| 1s | 0.750 | 1s | 1.627 | 1s | 2.789 |
| 2p | 4.000 | 1s | 1.105 | 1s | 1.992 |
| 2p | 2.051 | 1s | 0.750 | 1s | 1.423 |
| 2p | 1.052 | 2p | 7.667 | 1s | 1.016 |
| 3d | 3.400 | 2p | 3.932 | 2p | 11.667 |
| 3d | 1.900 | 2p | 2.016 | 2p | 6.306 |
| | | 2p | 1.034 | 2p | 3.409 |
| | | 3d | 5.000 | 2p | 1.843 |
| | | 3d | 3.125 | 2p | 0.996 |
| | | 3d | 1.953 | 3d | 5.900 |
| | | 4f | 2.600 | 3d | 4.214 |
| | | | | 3d | 3.010 |
| | | | | 3d | 2.150 |
| | | | | 4f | 3.600 |
| | | | | 4f | 1.800 |
| **N** | | | | | |
| 1s | 10.820 | 1s | 14.453 | 1s | 12.393 |
| 1s | 6.345 | 1s | 9.471 | 1s | 8.728 |
| 1s | 3.720 | 1s | 6.206 | 1s | 6.147 |
| 1s | 2.181 | 1s | 4.067 | 1s | 4.329 |
| 1s | 1.279 | 1s | 2.665 | 1s | 3.049 |
| 1s | 0.750 | 1s | 1.747 | 1s | 2.147 |
| 2p | 4.400 | 1s | 1.145 | 1s | 1.512 |
| 2p | 2.588 | 1s | 0.750 | 1s | 1.065 |
| 2p | 1.522 | 2p | 6.533 | 1s | 0.750 |
| 3d | 4.000 | 2p | 3.960 | 2p | 4.933 |
| 3d | 2.000 | 2p | 2.400 | 2p | 3.524 |
| | | 2p | 1.454 | 2p | 2.517 |
| | | 3d | 5.500 | 2p | 1.798 |

|    |        | 3d | 3.235 | 2p | 1.284 |
|----|--------|----|-------|----|-------|
|    |        | 3d | 1.903 | 3d | 5.800 |
|    |        | 4f | 2.200 | 3d | 4.143 |
|    |        |    |       | 3d | 2.959 |
|    |        |    |       | 3d | 2.114 |
|    |        |    |       | 4f | 4.200 |
|    |        |    |       | 4f | 2.100 |
| colspan O ||||||
| 1s | 13.351 | 1s | 18.099 | 1s | 14.948 |
| 1s | 7.506  | 1s | 11.485 | 1s | 10.283 |
| 1s | 4.220  | 1s | 7.288  | 1s | 7.075  |
| 1s | 2.373  | 1s | 4.625  | 1s | 4.867  |
| 1s | 1.334  | 1s | 2.935  | 1s | 3.348  |
| 1s | 0.750  | 1s | 1.862  | 1s | 2.303  |
| 2p | 5.700  | 1s | 1.182  | 1s | 1.585  |
| 2p | 2.850  | 1s | 0.750  | 1s | 1.090  |
| 2p | 1.425  | 2p | 7.000  | 1s | 0.750  |
| 3d | 4.000  | 2p | 4.000  | 2p | 8.667  |
| 3d | 2.000  | 2p | 2.286  | 2p | 5.417  |
|    |        | 2p | 1.306  | 2p | 3.385  |
|    |        | 3d | 6.000  | 2p | 2.116  |
|    |        | 3d | 3.529  | 2p | 1.322  |
|    |        | 3d | 2.076  | 3d | 5.600  |
|    |        | 4f | 2.600  | 3d | 4.000  |
|    |        |    |        | 3d | 2.857  |
|    |        |    |        | 3d | 2.041  |
|    |        |    |        | 4f | 4.400  |
|    |        |    |        | 4f | 2.200  |
| colspan F ||||||
| 1s | 15.520 | 1s | 21.908 | 1s | 17.972 |
| 1s | 8.466  | 1s | 13.528 | 1s | 12.082 |
| 1s | 4.619  | 1s | 8.354  | 1s | 8.123  |
| 1s | 2.520  | 1s | 5.158  | 1s | 5.461  |
| 1s | 1.375  | 1s | 3.185  | 1s | 3.671  |
| 1s | 0.750  | 1s | 1.967  | 1s | 2.468  |
| 2p | 5.667  | 1s | 1.215  | 1s | 1.659  |
| 2p | 3.063  | 1s | 0.750  | 1s | 1.116  |
| 2p | 1.656  | 2p | 5.667  | 1s | 0.750  |
| 3d | 4.000  | 2p | 3.542  | 2p | 7.000  |
| 3d | 2.000  | 2p | 2.214  | 2p | 4.828  |
|    |        | 2p | 1.383  | 2p | 3.329  |
|    |        | 3d | 5.000  | 2p | 2.296  |
|    |        | 3d | 3.226  | 2p | 1.584  |
|    |        | 3d | 2.081  | 3d | 6.500  |
|    |        | 4f | 2.300  | 3d | 4.333  |
|    |        |    |        | 3d | 2.889  |
|    |        |    |        | 3d | 1.926  |
|    |        |    |        | 4f | 5.400  |
|    |        |    |        | 4f | 2.700  |
| colspan Si ||||||
| 1s | 19.317 | 1s | 31.922 | 1s | 31.922 |
| 1s | 12.870 | 1s | 22.699 | 1s | 22.699 |
| 1s | 8.575  | 1s | 16.140 | 1s | 16.140 |
| 1s | 5.713 | 1s | 11.477 | 1s | 11.477 |
| 1s | 3.806 | 1s | 8.161 | 1s | 8.161 |
| 1s | 2.536 | 1s | 5.803 | 1s | 5.803 |
| 1s | 1.690 | 1s | 4.126 | 1s | 4.126 |
| 1s | 1.126 | 1s | 2.934 | 1s | 2.934 |
| 1s | 0.750 | 1s | 2.086 | 1s | 2.086 |
| 2p | 11.333 | 1s | 1.483 | 1s | 1.483 |
| 2p | 6.869 | 1s | 1.055 | 1s | 1.055 |
| 2p | 4.163 | 1s | 0.750 | 1s | 0.750 |
| 2p | 2.523 | 2p | 10.667 | 2p | 13.600 |
| 2p | 1.529 | 2p | 7.619 | 2p | 9.714 |
| 2p | 0.927 | 2p | 5.442 | 2p | 6.939 |
| 3d | 3.400 | 2p | 3.887 | 2p | 4.956 |
| 3d | 2.000 | 2p | 2.777 | 2p | 3.540 |
| | | 2p | 1.983 | 2p | 2.529 |
| | | 2p | 1.417 | 2p | 1.806 |
| | | 2p | 1.012 | 2p | 1.290 |
| | | 3d | 3.500 | 2p | 0.922 |
| | | 3d | 2.500 | 2p | 0.658 |
| | | 3d | 1.786 | 3d | 5.400 |
| | | 4f | 2.200 | 3d | 3.857 |
| | | | | 3d | 2.755 |
| | | | | 3d | 1.968 |
| | | | | 4f | 4.400 |
| | | | | 4f | 2.200 |
| | | S | | | |
| 1s | 23.379 | 1s | 31.547 | 1s | 31.547 |
| 1s | 15.209 | 1s | 22.456 | 1s | 22.456 |
| 1s | 9.894 | 1s | 15.985 | 1s | 15.985 |
| 1s | 6.437 | 1s | 11.380 | 1s | 11.378 |
| 1s | 4.187 | 1s | 8.099 | 1s | 8.099 |
| 1s | 2.724 | 1s | 5.765 | 1s | 5.765 |
| 1s | 1.772 | 1s | 4.104 | 1s | 4.104 |
| 1s | 1.153 | 1s | 2.921 | 1s | 2.921 |
| 1s | 0.750 | 1s | 2.079 | 1s | 2.079 |
| 2p | 12.800 | 1s | 1.480 | 1s | 1.480 |
| 2p | 8.000 | 1s | 1.054 | 1s | 1.054 |
| 2p | 5.000 | 1s | 0.750 | 1s | 0.750 |
| 2p | 3.125 | 2p | 13.067 | 2p | 17.067 |
| 2p | 1.953 | 2p | 9.333 | 2p | 12.190 |
| 2p | 1.221 | 2p | 6.667 | 2p | 8.707 |
| 3d | 3.700 | 2p | 4.762 | 2p | 6.220 |
| 3d | 2.000 | 2p | 3.401 | 2p | 4.443 |
| | | 2p | 2.430 | 2p | 3.173 |
| | | 2p | 1.735 | 2p | 2.267 |
| | | 2p | 1.240 | 2p | 1.619 |
| | | 3d | 3.500 | 2p | 1.156 |
| | | 3d | 2.500 | 2p | 0.826 |
| | | 3d | 1.786 | 3d | 4.600 |
| | | 4f | 2.300 | 3d | 3.286 |
| | | | | 3d | 2.347 |
| | | | | 3d | 1.676 |
| | | | | 4f | 2.800 |

|     |        |     |        | 4f  | 1.400  |
|-----|--------|-----|--------|-----|--------|
|     |        |     | Cl     |     |        |
| 1s  | 21.825 | 1s  | 31.403 | 1s  | 31.747 |
| 1s  | 15.156 | 1s  | 22.363 | 1s  | 22.585 |
| 1s  | 10.525 | 1s  | 15.925 | 1s  | 16.068 |
| 1s  | 7.309  | 1s  | 11.341 | 1s  | 11.431 |
| 1s  | 5.076  | 1s  | 8.076  | 1s  | 8.132  |
| 1s  | 3.525  | 1s  | 5.751  | 1s  | 5.785  |
| 1s  | 2.448  | 1s  | 4.095  | 1s  | 4.116  |
| 1s  | 1.700  | 1s  | 2.916  | 1s  | 2.928  |
| 1s  | 1.180  | 1s  | 2.077  | 1s  | 2.083  |
| 2p  | 22.000 | 1s  | 1.479  | 1s  | 1.482  |
| 2p  | 12.571 | 1s  | 1.053  | 1s  | 1.054  |
| 2p  | 7.184  | 1s  | 0.750  | 1s  | 0.750  |
| 2p  | 4.105  | 2p  | 14.667 | 2p  | 18.600 |
| 2p  | 2.346  | 2p  | 10.476 | 2p  | 13.286 |
| 2p  | 1.340  | 2p  | 7.483  | 2p  | 9.490  |
| 3d  | 3.7    | 2p  | 5.345  | 2p  | 6.778  |
| 3d  | 2.0    | 2p  | 3.818  | 2p  | 4.842  |
|     |        | 2p  | 2.727  | 2p  | 3.458  |
|     |        | 2p  | 1.948  | 2p  | 2.470  |
|     |        | 2p  | 1.391  | 2p  | 1.764  |
|     |        | 3d  | 4.000  | 2p  | 1.260  |
|     |        | 3d  | 2.857  | 2p  | 0.900  |
|     |        | 3d  | 2.041  | 3d  | 5.100  |
|     |        | 4f  | 2.200  | 3d  | 3.643  |
|     |        |     |        | 3d  | 2.602  |
|     |        |     |        | 3d  | 1.859  |
|     |        |     |        | 4f  | 2.400  |
|     |        |     |        | 4f  | 1.200  |

## II. Geometries

**Table S2:** Geometries in Ångstrom for systems used in Figures 3, 4, 5, 7 and Tables 1, 2, and 3 as well as the large bond distance $OF_2$ structure used to obtain Figure S1. Geometries with 3 or greater atoms were optimized using DFT with a ωB97X functional in a cc-pVQZ basis, unless specified elsewhere. The 1,3 propanediyl system was optimized with the B3LYP functional in a 6-31G* basis.

| Molecule | Atom | X   | Y   | Z      |
|----------|------|-----|-----|--------|
| HF       |      |     |     |        |
|          | H    | 0.0 | 0.0 | 0.0    |
|          | F    | 0.0 | 0.0 | 0.9170 |
| $C_2$    |      |     |     |        |
|          | C    | 0.0 | 0.0 | 0.0    |
|          | C    | 0.0 | 0.0 | 1.147  |
| $N_2$    |      |     |     |        |
|          | N    | 0.0 | 0.0 | 0.0    |
|          | N    | 0.0 | 0.0 | 1.098  |
| CO       |      |     |     |        |
|          | C    | 0.0 | 0.0 | 0.0    |
|          | O    | 0.0 | 0.0 | 1.128  |
| NO       |      |     |     |        |
| | N | 0.0 | 0.0 | 0.0 |
| | O | 0.0 | 0.0 | 1.054 |
| $O_2$ | | | | |
| | O | 0.0 | 0.0 | 0.0 |
| | O | 0.0 | 0.0 | 1.208 |
| HCl | | | | |
| | H | 0.0 | 0.0 | 0.0 |
| | Cl | 0.0 | 0.0 | 1.275 |
| CS | | | | |
| | C | 0.0 | 0.0 | 0.0 |
| | S | 0.0 | 0.0 | 1.535 |
| SiO | | | | |
| | Si | 0.0 | 0.0 | 0.0 |
| | O | 0.0 | 0.0 | 1.510 |
| ClF | | | | |
| | Cl | 0.0 | 0.0 | 1.628 |
| | F | 0.0 | 0.0 | 0.0 |
| $Cl_2$ | | | | |
| | Cl | 0.0 | 0.0 | 0.0 |
| | Cl | 0.0 | 0.0 | 1.988 |
| $H_2O$ | | | | |
| | H | 0.03662 | 0.2867 | -0.7420 |
| | H | 0.006501 | 1.1120 | 0.5315 |
| | O | -0.003118 | 0.2113 | 0.2105 |
| HCN | | | | |
| | H | -0.008949 | 0.01992 | 1.092 |
| | C | 0.01279 | 0.05594 | 0.02497 |
| | N | 0.03616 | 0.09413 | -1.117 |
| OFH | | | | |
| | O | -0.004260 | 0.02793 | 0.1246 |
| | F | 0.005461 | 1.316 | 0.6854 |
| | H | 0.03880 | 0.2657 | -0.8100 |
| $N_3$ | | | | |
| | N | 0.01331 | 0.5362 | 0.0003591 |
| | N | -0.005540 | 1.255 | 0.9236 |
| | N | 0.03223 | -0.1809 | -0.9239 |
| $N_2O$ | | | | |
| | N | 0.03390 | -0.1616 | -1.002 |
| | N | 0.01371 | 0.3602 | -0.01847 |
| | O | -0.007608 | 0.9114 | 1.020 |
| $NO_2$ | | | | |
| | N | -0.008446 | -0.02460 | 0.1907 |
| | O | 0.003103 | 1.029 | 0.8568 |
| | O | 0.04534 | 0.1060 | -1.0475 |
| SFH | | | | |
| | S | -0.01224 | -0.07200 | 0.2630 |
| | F | 0.005247 | 1.462 | 0.7820 |
| | H | 0.04700 | 0.2196 | -1.045 |
| $OF_2$ | | | | |
| | O | -0.01081 | 0.06083 | 0.3100 |
| | F | 0.004032 | 1.367 | 0.7531 |
| | F | 0.04678 | 0.1819 | -1.063 |
| $SO_2$ | | | | |

|   |   |   | -0.01091 | -0.06565 | 0.2156 |
|---|---|---|---|---|---|
|   |   | S | 0.001496 | 1.134 | 0.9929 |
|   |   | O | 0.04941 | 0.04196 | -1.208 |
|   | CS$_2$ | O |   |   |   |
|   |   | C | 0.01333 | 0.5366 | 0.00005083 |
|   |   | S | -0.01564 | 1.410 | 1.275 |
|   |   | S | 0.04231 | -0.3368 | -1.275 |
|   | OF$_2$ (long bond) |   |   |   |   |
|   |   | O | 0.0 | 0.0 | 0.0 |
|   |   | F | 0.0 | 0.0 | 1.380 |
|   |   | F | 0.0 | 40.23 | -9.810 |
|   | 1,3 Propanediyl |   |   |   |   |
|   |   | C | 1.278 | 0.0 | 0.2597 |
|   |   | C | 0.0 | 0.0 | -0.5852 |
|   |   | C | -1.278 | 0.0 | 0.2597 |
|   |   | H | 1.325 | -0.8846 | 0.9073 |
|   |   | H | 1.325 | 0.8846 | 0.9073 |
|   |   | H | 0.0 | 0.8776 | -1.246 |
|   |   | H | 0.0 | -0.8776 | -1.246 |
|   |   | H | -1.325 | -0.8846 | 0.9073 |
|   |   | H | -1.325 | 0.8846 | 0.9073 |

<br>

### III. PS Grid Sizes

**Table S3:** Grid sizes used for Figures 3 and 4 (left) and Figure 5 (right). All combinations of radial and angular number of points were considered (left). Each row of values represents parameters for a single calculation in the generation of Figure 5 (right).

| $N_\mu$ | $N_\nu$ | $N_\phi$ |
|---|---|---|
| 26 | 20 | 8 |
| 27 | 22 | 9 |
| 28 | 24 | 10 |
| 29 | 26 | 11 |
| 30 | 28 | 12 |
| 31 | 30 | 13 |
| 32 | 32 | 14 |
| 33 | 34 | 15 |
| 34 | 36 | 16 |
| 35 | 38 | 17 |
| 36 | 40 | 18 |
| 37 | 42 | 19 |

| $N_\mu$ | $N_\nu$ | $N_\phi$ |
|---|---|---|
| 13 | 20 | 8 |
| 14 | 20 | 8 |
| 15 | 22 | 9 |
| 16 | 22 | 9 |
| 17 | 24 | 10 |
| 18 | 24 | 10 |
| 19 | 26 | 11 |
| 20 | 26 | 11 |
| 21 | 28 | 12 |
| 22 | 28 | 12 |
| 23 | 30 | 13 |
| 24 | 30 | 13 |
| 25 | 32 | 14 |
| 26 | 32 | 14 |
| 27 | 34 | 15 |
| 28 | 34 | 15 |
| 29 | 36 | 16 |
| 30 | 36 | 16 |
| 31 | 38 | 17 |
| 32 | 38 | 17 |
| 33 | 40 | 18 |
| 34 | 40 | 18 |
| 35 | 42 | 19 |

|  |  |  | 36 | 42 | 19 |
|---|---|---|---|---|---|
|  |  |  | 37 | 44 | 20 |
| 38 | 44 | 20 | 38 | 44 | 20 |

## IV. Figure S1

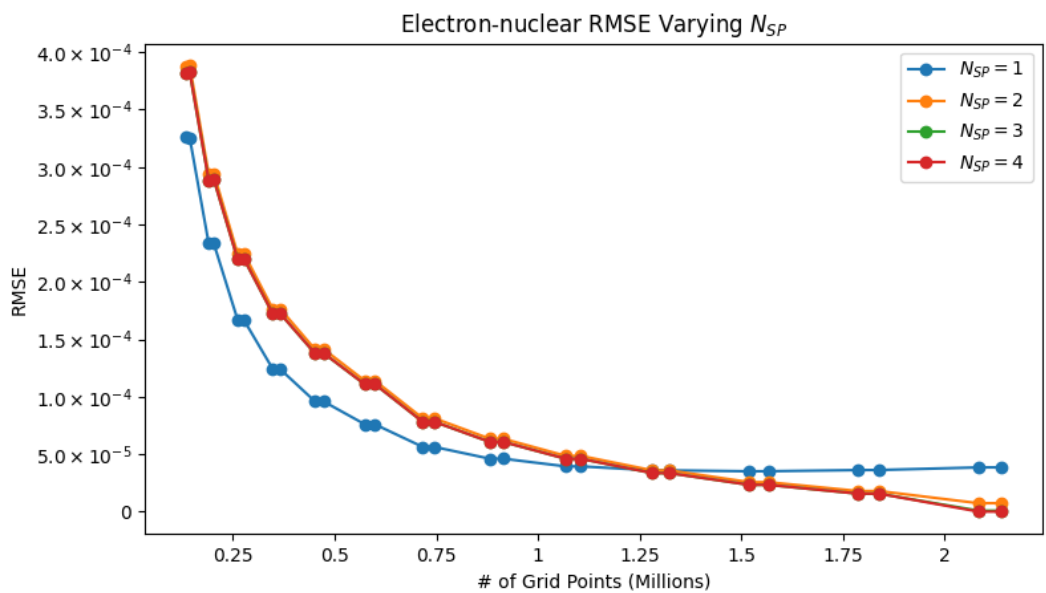

**Figure S1:** Comparison of the RMSE (Ha) of the electron-nuclear attraction elements for different divisions of the third center within the PS integration grid. Calculations performed on the $OF_2$ system with one OF bond at a 41.41Å distance.

    The RMSE of the electron-nuclear attraction integral for $OF_2$ with an OF bond at ~30 times equilibrium bond length confirms the error is reduced at large grid discretization up to $N_{SP} = 3$. At smaller grid sizes when $N_{SP} > 1$ the RMSE of the electron-nuclear attraction integral increases, this phenomenon is believed to be caused by a cancellation of error when the grid around the third center is coarser.

## V. Figure S2

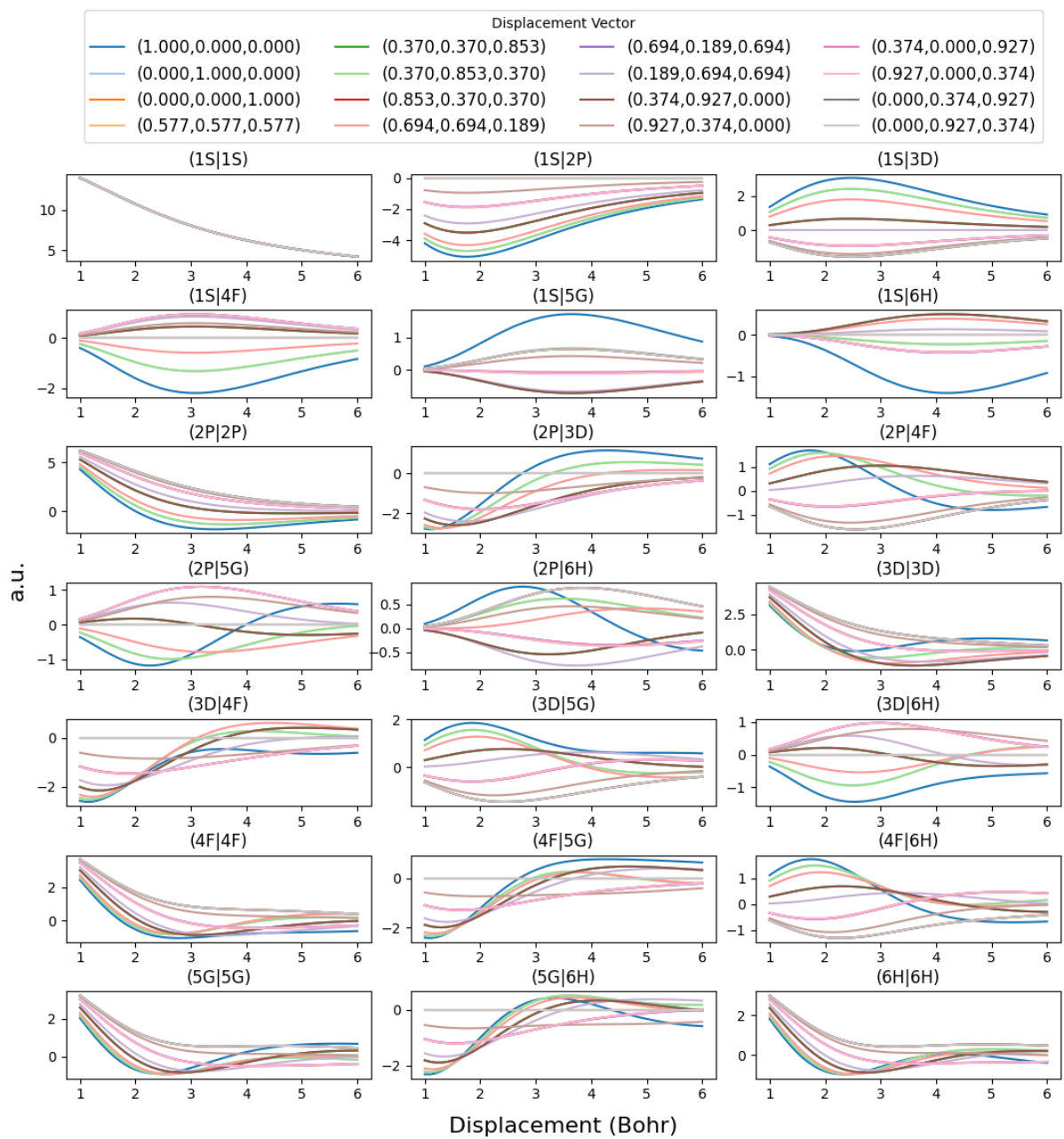

**Figure S2:** All combinations of 2-center Coulomb integrals plotted along 16 vector directions[15] as the two centers are separated. The legend indicates unit vectors for all 16 directions tested, all basis functions used have the exponent $\zeta = 1$, and all basis functions were chosen to have $m = 0$.

## VI. Figures S3 and S4

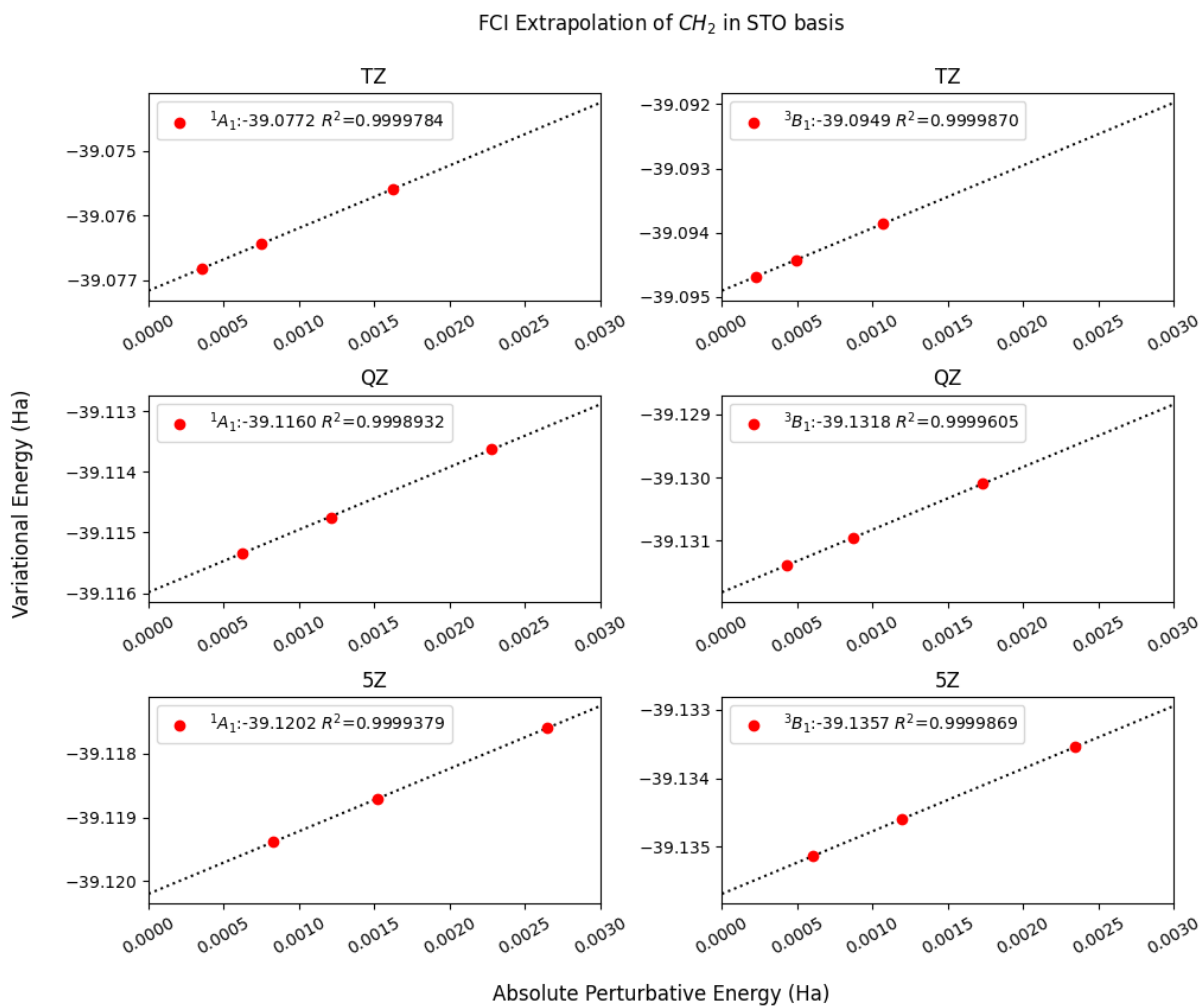

**Figure S3:** FCI extrapolation energies for methylene in the TZ, QZ, and 5Z basis for both the $^1A_1$ and $^3B_1$ states using STOs.

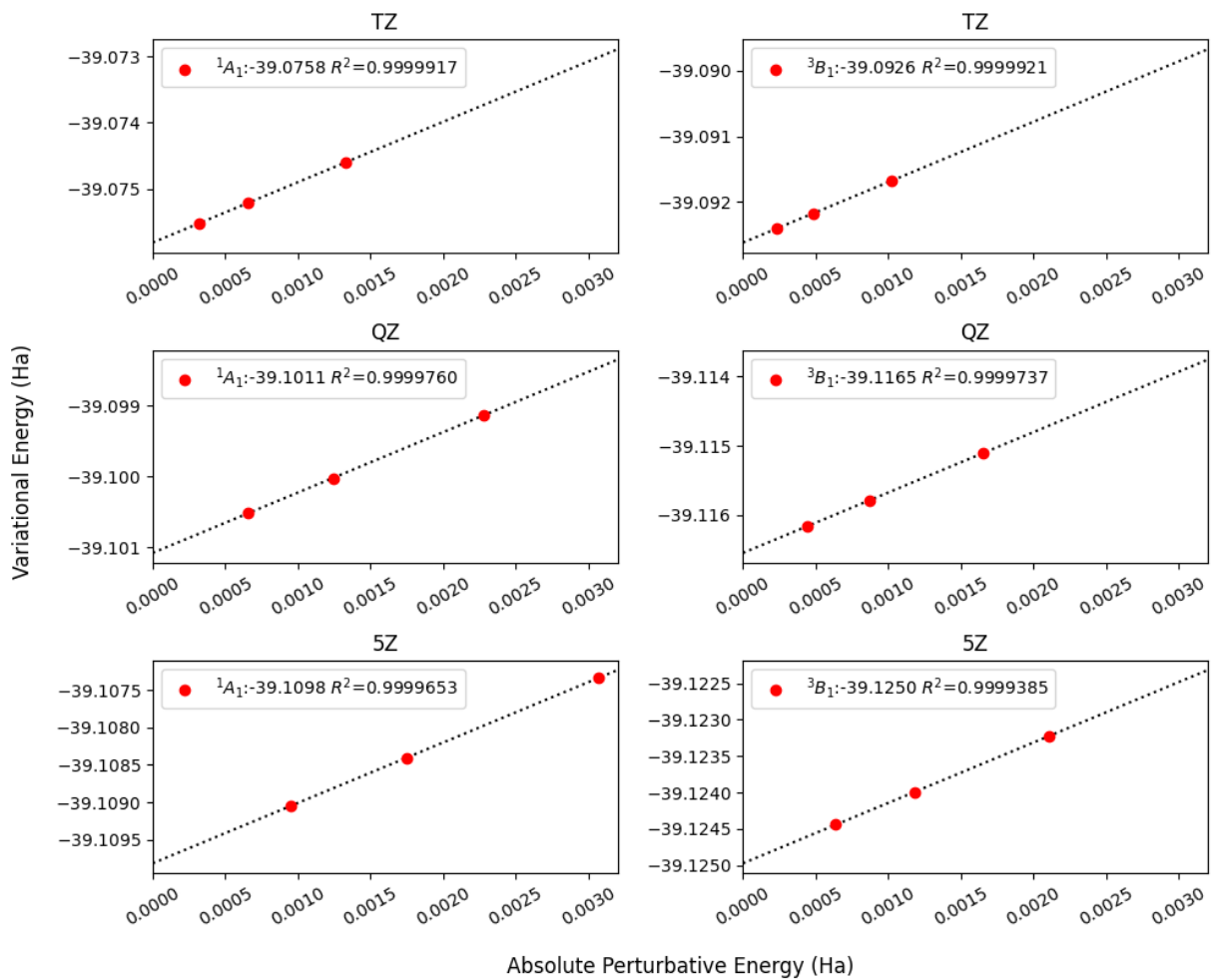

**Figure S4:** FCI extrapolation energies for methylene in the cc-pVTZ (TZ), cc-pVQZ (QZ), and cc-pV5Z (5Z) basis for both the $^1A_1$ and $^3B_1$ states using GTOs.

To obtain singlet-triplet gap energies for methylene with basis XZ (where X = T, Q, 5) three calculations are performed with varying $\varepsilon_1$ values ($\varepsilon_1 = 2 \times 10^{-4}, 1 \times 10^{-4}, 0.5 \times 10^{-4}$ Ha and $\varepsilon_2 = 5 \times 10^{-7}$ Ha) for the singlet and the triplet systems. The energies obtained from the extrapolations are then subtracted to find the singlet-triplet gap.